# Uncharacteristic second order martensitic transformation in metals via epitaxial stress fields

Authors: Samuel Temple Reeve[1], Karthik Guda Vishnu[2], and Alejandro Strachan[2*]

Affiliations:

[1]Materials Science Division, Lawrence Livermore National Laboratory,
7000 East Ave, Livermore, CA 94550 USA

[2]School of Materials Engineering and Birck Nanotechnology Center, Purdue University,
West Lafayette, Indiana 47906 USA

*Correspondence to: strachan@purdue.edu

Abstract

While most phase transformations, e.g. ferroelectric or ferromagnetic, can be first or second order depending on external applied fields, martensitic transformations in metallic alloys are nearly universally first order. We demonstrate that epitaxial stress originating from the incorporation of a tailored second phase can modify the free energy landscape that governs the phase transition and change its order from first to second. High-fidelity molecular dynamics simulations show a remarkable change in the character of the martensitic transformation in Ni-Al alloys near the critical point. We observe the continuous evolution of the transformation order parameter and scaling with power-law exponents comparable to those in other ferroic transitions exhibiting critical behavior. Our theoretical work provides a foundation to recent experimental and computational results on martensites near critical points.

## I. Introduction

Phase transformations are fascinating from a scientific point of view and critical for human development, from the production of early bronzes through the rise of steels[1] to today's phase change materials used in solar energy[2] and nanoelectronics[3]. In the field of materials science, ferroelectric transformations power actuators and sensors[4,5] and enhance the performance of electronic devices[6]. Analogously, solid-to-solid martensitic transformations in metallic alloys, underlie shape memory, superelasticity, and strengthening in many high-performance steels[7,8]. Under most conditions, phase transformations involve a discontinuous jump in properties; for example, as a liquid turns into vapor its density changes abruptly and as a ferroelectric material is cooled down across its Curie temperature, polarization develops suddenly[9]. However, in most cases, the character of the transformation can be changed from discontinuous (classified as first order) to continuous (second order) via the application of an external field conjugate with the order parameter that governs the transformation: electric and magnetic fields in ferroelectrics[10] and ferromagnetics, respectively, and pressure in the liquid-gas transition[9]. The change from first to second order is accompanied by striking effects, from the existence of universal scaling laws to a continuous transition lacking hysteresis. Scaling laws and the existence of universality classes, i.e. when properly scaled the behavior of various systems collapsing into identical behavior, imply shared underlying physics among disparate systems and phenomena. From an engineering perspective, the continuous nature and lack of hysteresis can be harnessed in various applications such as sensors and actuators[11,12]. These applications would benefit from a lack of abrupt shape changes that often lead to functional fatigue, as well as a reduction in the hysteretic losses during a full cycle. While the existence of a critical point associated with a second order transformation is common to most phase changes, martensitic transformations in metallic alloys are stubbornly first order. While specific compositions can lead to continuous behavior[13], particularly in systems with small changes in lattice parameter and crystal structure, e.g. FCC-FCT[14–16], only one study is known to the authors in which an external field, mechanical stress, was used to continuously tune the nature of these transformations and achieve criticality[17]. In this work we go a step further, using the internal strain fields resulting from a tailored coherent second phase to transform the nature of martensitic transformation in metallic alloys from first to second order and provide further characterization of their criticality. With theory and high-fidelity molecular dynamics (MD) simulations, we therefore demonstrate the possibility of material level control of the transformation order, without requiring an external load.

A key characteristic of second order phase transformations is that many accompanying properties are described by scaling laws reflecting a lack of characteristic scales. More importantly, disparate systems often exhibit identical scaling exponents, indicating underlying universal behavior. For example, ferroic transitions, order-disorder transitions, superconductors, and superfuids[9] belong to the same universality class. Many systems beyond this list of classic examples also display critical behavior, from biology[18], to Mott transitions[19], liquid crystals[20,21], and granular materials[22]. Our MD simulations on NiAl alloys confirm this dramatic change in the nature of transformation in the vicinity of the critical point, showing scaling behavior with exponents consistent with the mean field universality class. Our findings provide a theoretical foundation for the recent experiments with the aforementioned Fe-Pd alloy, in Ti-based gum metals containing nanoscale variations in composition and exhibiting "higher-order" behavior including continuous stress-induced transformation[23,24], as well as computational prediction of ultra-low stiffness[25], reduced hysteresis, and ability to tune transformation temperatures[26,27] in Ni-Al alloys.

## II. Methods and analysis

Quenched $Ni_xAl_{1-x}$ alloys display a martensitic transformation and shape memory between 60 and 65 at. % Ni[28]. Here, off-stoichiometric 63 at. % Ni – 37 at. % Al is the martensitic phase and B2 50 at. % Ni – 50 at. % Al (NiAl)the non-martensitic second phase. The interatomic interactions were described with a potential developed by Farkas et al.[29], fit to experimental properties including cohesive energies, lattice parameters, and elastic constants of Ni, Al, $Ni_3Al$, NiAl, $Ni_5Al_3$, and $L1_0$ martensite. It has been shown to accurately capture a cubic to monoclinic martensitic transformation with reasonable transformation temperatures for the correct range of Ni compositions[26]. The model and motivation of the approach are described in more detail in prior publications[25,26,30,31]. We stress that our objective is to show how a second phase can affect the nature of a martensitic transformation, thus, the accuracy of the potential in describing the specifics of NiAl is of secondary importance. However, in our prior work we have shown that two independent potentials for NiAl resulted in overall similar trends regarding the effect of interfacial stresses on mechanical properties[25]. Given this test and description of the underlying physics of the system, we believe our results are not dependent on the specific choice of interatomic potential.

All MD simulations were performed using LAMMPS[32]. All bulk and nanolaminate systems were created with 1,024,000 atoms, with 80 B2 unit cells in each direction, initially a cube measuring 23.2 nm. Throughout, 3D periodic boundary conditions were used. Regions of off-stoichiometric 63 at. % Ni were created by randomly swapping Ni on Al sites within a layer of the desired volume fraction. These systems match those from Ref. 27 with additional volume fractions above 60 at. % NiAl. A timestep of 1 fs with velocity Verlet integration was used throughout, as well as Nosé-Hoover thermostat and barostat coupling constants of 0.01 ps and 0.1 ps, respectively. The relaxation timescales with these damping parameters are at least an order of magnitude smaller than the fastest transformation timescales and several orders of magnitude faster than the nanolaminate cases of interest.

All visualization and cluster analysis was performed with OVITO[33]. The polyhedral template matching (PTM) algorithm[34] implemented in OVITO was used for structural identification. For each atom, the local neighbor structure is compared to templates for common crystals (FCC, BCC, HCP, etc.) and the root mean square (RMS) error calculated. Each atom is then identified as the structure with the lowest error. For this work, atoms identified as BCC are referred to as austenite, HCP as martensite, and FCC as stacking faults. If no mapping exists to any of the chosen templates, the atom is listed as "other". In addition, an RMS error cutoff of 0.12 was chosen, above which atoms were also considered "other".

Martensite clusters were analyzed by deleting all austenite atoms with OVITO cluster analysis to determine number of clusters and OVITO surface mesh analysis to extract cluster surface area and volume for each system and temperature.

## III. Engineering martensitic free energy landscapes into a continuous transformation

The discontinuous nature of a first order phase transition originates from the shape of the free energy landscapes underlying the transformation and their temperature dependence. Figure 1A shows the free energy as a function of lattice parameter for a model $Ni_{63}Al_{37}$ alloy obtained using MD for various temperatures. Free energy landscapes were calculated from the differential expression of Helmholtz free energy starting from the stable, unstrained state for each temperature and strained at a rate of $1 \cdot 10^9$ ps$^{-1}$ in tension and compression. Under isothermal conditions, the entropic term does not contribute to the differential of free energy and changes in free energy can be obtained by integrating the following:

$$dF = V(\sigma_{xx}d\varepsilon_{xx} + \sigma_{yy}d\varepsilon_{yy} + \sigma_{zz}d\varepsilon_{zz} + \tau_{xy}d\gamma_{xy} + \tau_{xz}d\gamma_{xz} + \tau_{yz}d\gamma_{yz}) \quad (1)$$

where $F$ is the free energy, $V$ the system volume, $\sigma$ and $\tau$ are stress components, and $\varepsilon$ and $\gamma$ are strain components. Details of the stress calculations in LAMMPS are given in Ref. 35.

From the landscapes in Fig. 1 we see that at high temperatures the austenite phase (A) is stable and the martensite (M) is metastable. Both phases have approximately the same free energy at 950K and at lower temperatures the martensite phase is stable. These landscapes represent the typical behavior of first order transformations. Note that the two phases are separated by an energy barrier that must be overcome during the phase transformation (resulting in hysteresis) and that the austenite and martensite phases have distinct lattice parameters at all temperatures. Thus, the transformation involves a jump in properties. We hypothesized that a continuous change from one phase to another, without an energy barrier, could be achieved by modifying this martensitic energy landscape via a coherent second phase with a complementary landscape. Coherent integration at the nanoscale, involving defect free interfaces as shown in Fig. 1D, forces the two phases to share the same lattice parameters; consequently, one can think of the energy landscape of the composite metamaterial as the weighted sum of each component as a function of lattice parameter. Figure 1B exemplifies our approach of free energy landscape engineering (FELE) to achieve a barrier-less transformation. Building on our previous work[25–27], we considered phases with opposing stability: $Ni_{63}Al_{37}$ far below austenite metastability and the ordered $Ni_{50}Al_{50}$ for which the martensite is unstable down to 0 K. An analytical combination of these two landscapes, at 25 K, yields the desired landscapes (dashed line in Fig. 1B) with a flat landscape characteristic of second order systems. While subtler than the analytical combination, the actual landscapes obtained from direct simulation of a coherent nanolaminate of 65% NiAl and 35% $Ni_{63}Al_{37}$, Fig. 1C, indeed shows typical features expected in a second order phase transformation. The system continuously moves from the high-temperature austenite to the martensite phase without a barrier.

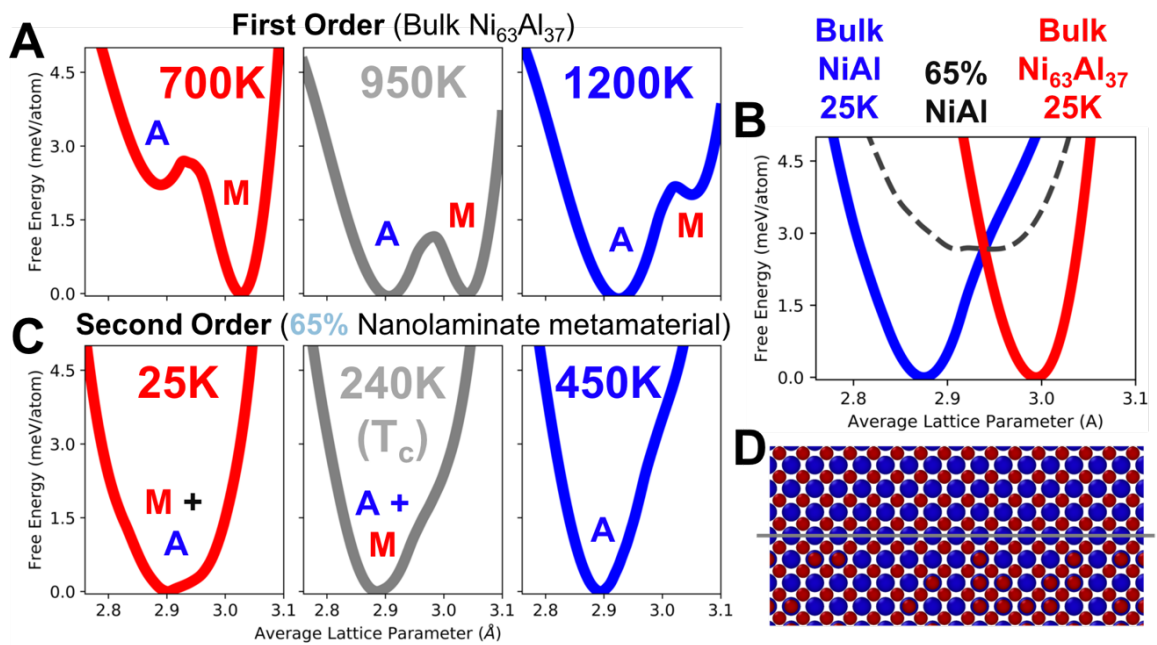

**Fig. 1.** Free energy landscape engineering producing the possibility of second order behavior. (A) Calculated free energy landscapes for a bulk, first order $Ni_{63}Al_{37}$ martensitic alloy, (B) analytically combined free energy landscape from separate NiAl and $Ni_{63}Al_{37}$ bulk systems at low temperature, (C) free energy landscapes for a 65 at. % NiAl nanolaminate, and (D) atomic interface between NiAl (top) and $Ni_{63}Al_{37}$ (bottom) with Al blue and Ni red.

## IV. Critical martensitic transformation and scaling laws

The free energy landscape shown in Fig. 1C provides strong evidence for the second order nature of the transformation in the coherent metamaterial. To confirm this expectation and to characterize features of the transformation, we carried out explicit MD simulations of the thermal transformation. Coherent laminates with period 11.6 nm and cross-sectional lengths of 23.2 nm (interface shown in Fig. 1D) were cooled from above the martensitic transformation temperature ($M_s$) to 25 K under isothermal-isobaric conditions (NPT) at a rate of $1 \cdot 10^{11}$ K/s and then heated above the austenite finish ($A_f$) temperature at the same rate. During heating and cooling, all directions and angles were left free to relax at 1 atm pressure We characterized the nature of the thermal transformation by analyzing the thermodynamic response of the coherent metamaterial as a function of the volume fraction of the NiAl second phase and exploring possible scaling laws. In addition, a detailed local structure analysis of the atomistic trajectories was used to identify phases, revealing domain structure and size distributions characteristic of critical phenomena with marked differences from the bulk $Ni_{63}Al_{37}$ alloy.

## A. Strain order parameter exponent

We define the transformation order parameter as the martensitic transformation strain along the in-plane direction for the nanolaminates; this is analogous to ferroelectric polarization and ferromagnetic magnetization. The bulk alloy shows the baseline characteristics of transformation strain as a function of temperature (shifted by $M_s$), see the red line in Fig. 2. This transformation shows first order characteristics, with an abrupt jump from almost entirely austenite to fully martensite when $M_s$ is reached upon cooling. The addition of up to 50 at. % of the non-transforming NiAl phase does not change the transformation in a fundamental manner; we see an abrupt, but incomplete, transformation. However, as hypothesized from the free energy landscapes in Fig. 1C, 65 at. % NiAl and higher fractions results in distinct behavior, with a continuous increase in the order parameter. Despite the intrinsic fluctuations characteristic of relatively small systems with off-stoichiometric composition, the change from discontinuous to a continuous transformation is clear in Fig. 2. We note that the apparent thickness of the lines in Fig. 2 reflect the statistical fluctuations of the order parameter.

The critical exponent associated with the transformation order parameter (strain in this ferroelastic system) is known as $\beta$, and defined for the temperature dependence of the order parameter:

$$\varepsilon_T = (T - M_s)^{-\beta} \qquad (2)$$

where $\varepsilon_T$ is the transformation strain that represents the order parameter, $T$ is temperature, and $M_s$ is the martensite start temperature, more often referred to as the critical temperature ($T_c$) for critical behavior, and interchangeable in this context. As will be discussed in detail in Section IV C and in the Supplemental Material, the exponent $\beta$ obtained from our simulations ranges between 0.45 and 0.7 for 65 at. % NiAl, consistent with the values in similar critical phenomena. Further support of the critical nature of the transformation is a near zero hysteresis in the transformation for all systems above 62.5 at. % NiAl. Figure 3 shows the average lattice parameter of the supercell in the in-plane and out-of-plane directions during a complete cooling and heating cycle. We observe a monotonic reduction in hysteresis with increasing NiAl content.

In order to test the size dependence of the phenomena observed we performed additional simulations with system sizes of approximately 5 and 20 million atoms for 65 at. % NiAl to compare with 1 million atom systems described so far. Simulation size is scaled equally in all directions and details of the simulation cells is provided in the Supplemental Table 1. We find identical physics in these larger systems, with nearly identical behavior near the critical temperature. Supplemental Fig. 2 compares the temperature dependence of the order parameter and exponent $\beta$ for the larger systems with the behavior to the 1 million atom system. The most important difference we observe in the larger systems is a stronger effect from domain orientation compatibility leading to additional steps in the strain response, investigation of which is left to future work. These effects appear for temperatures away from the critical value.

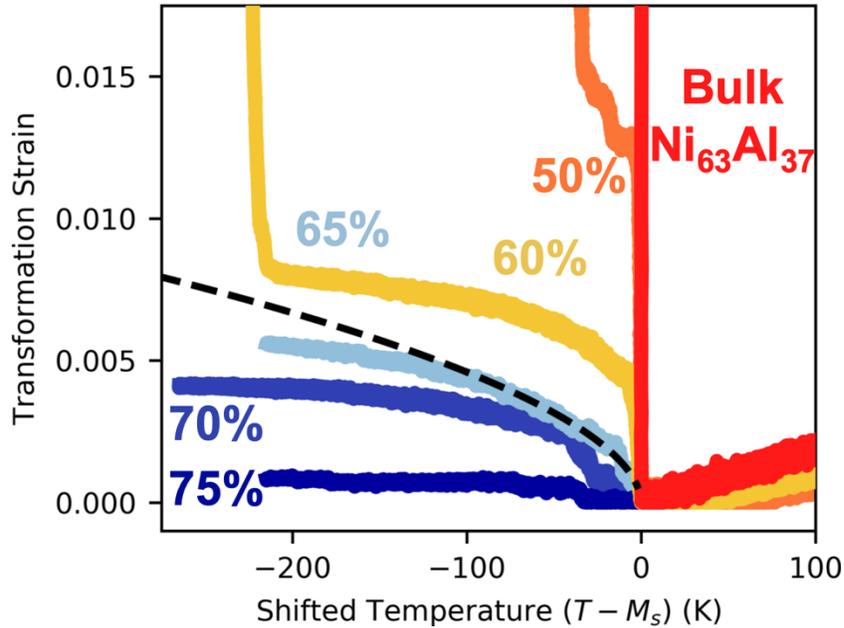

**Fig. 2.** Martensitic transformation strain order parameter. Bulk $Ni_{63}Al_{37}$ behavior contrasts the high NiAl volume fraction nanolaminates, all cooled to 25K. Fitting of scaling exponent shown for 65 at. % NiAl with Eq. 2.

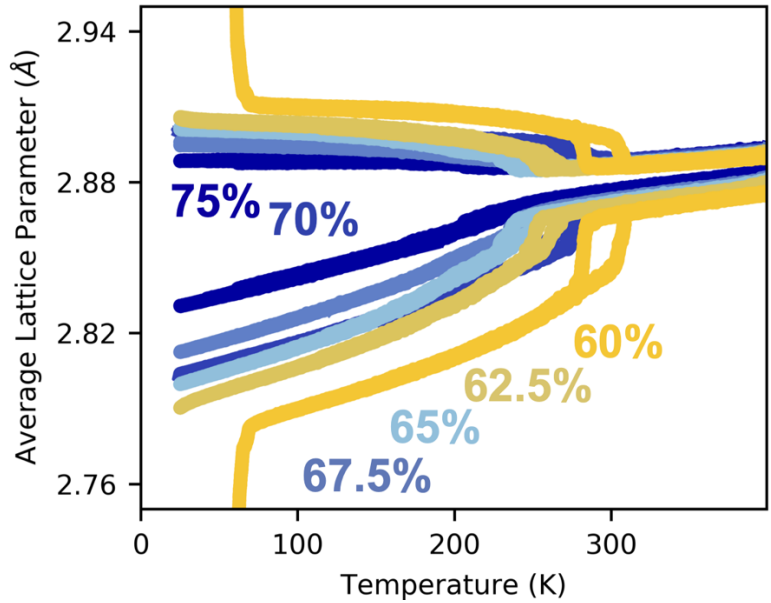

**Figure 3:** Cooling and heating cycle for all nanolaminate systems above 60 at. % NiAl. Note the lack of hysteresis for all nanolaminates greater than 62.5 at. % NiAl.

B. Scaling law describing domain structure

An analysis of the atomic structure during the phase transition provides important additional insight regarding the nature of the phase transition. Figure 4 shows atomistic structures for the 65 at. % NiAl metamaterial at two temperatures below $M_s$ (30 and 190 K). Atoms in the martensitic phase are colored in red and those in the austenite phase are transparent blue. The simulations reveal a fascinating domain structure with interpenetrating phases and rough interfaces, uncharacteristic for martensitic transformations which exhibit planar domain walls with preferred lattice matching between the austenite and martensite: $[100]_A \parallel [110]_M$ in this system. While martensitic domain walls and interfaces can have irregular structure close to nucleation due to compatibility with other domains or phases (e.g. austenite), overall the interfaces favor preferred planes, particularly after the nucleation[36]. In contrast, the domain structure in Fig. 4A is typical of second order transformations, where the difference between phases disappears leading to rough, jagged interfaces and a lack of a characteristic microstructural length under equilibrium conditions. Even as combinations of self-accommodating martensite domains form with continued cooling for overall compatibility, the rough, interpenetrating austenite/martensite interfaces persist (see Fig. 4B) without significant local orientation preference. We note that the transformation only occurs within the $Ni_{63}Al_{37}$ region of the nanolaminate and that the structure shown in Fig. 4 is representative of observations for NiAl at. % between 60 and 75, beyond which the transformation is completely suppressed.

To quantify these microstructures, we note that second order phase transformations are characterized by power law scaling in domain size distribution with critical exponent $\tau$:

$$N \propto V^{-\tau} \qquad (3)$$

where $N$ is the number of clusters of a given volume $V$. We performed a cluster analysis on the atoms in the martensitic phase to identify individual domains, see details in Supplemental Fig. 3. We find the cluster size distribution indeed follows a power law right below $M_s$ and until domains coalesce leading to one dominant martensitic domain, shown in Fig. 5 as an unconnected point. We find identical scaling in the larger simulation cells (with 5 and 20 million atoms) and nearly identical exponents $\tau$, see Supplemental Fig. 4.

Supplemental Video 1 shows this cluster development throughout the cooling process, starkly contrasted by the bulk $Ni_{63}Al_{37}$ system in Supplemental Video 2. Finally, while the surface area and volume of 3D martensitic domains in first order transformations scale as 2/3 ($A = V^{2/3}$), these martensite clusters in this second order transformation scale with an exponent of 0.86, or a fractal dimensionality of 2.58 (Supplemental Fig. 5). Resulting values of $\tau$ and fractal dimension are consistent across volume fraction until complete transformation suppression.

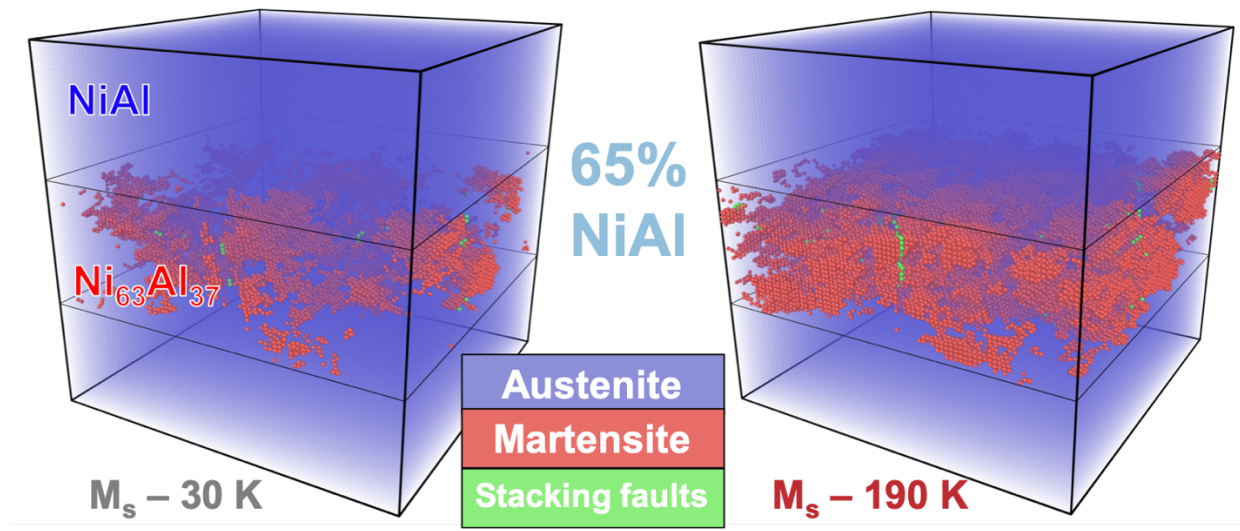

**Fig. 4.** Cluster fractal characteristics. Atomistic structure of the 65 at. % NiAl, exemplifying the localized, continuous nature of the martensitic transformation at two temperatures, 30 K and 190 K below $M_s$. Atoms classified as austenite are shown as transparent blue, martensite as red, and stacking faults as green.

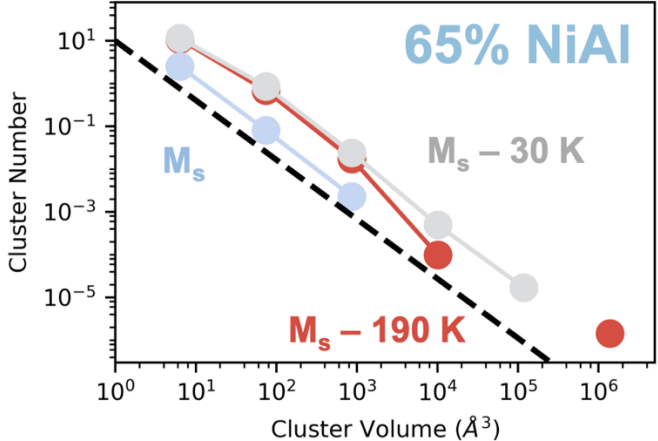

**Fig. 5.** Cluster scaling behavior. Cluster distribution showing scaling at $M_s$, extended scaling slightly below at $M_s$ - 30 K, and one large, coalesced cluster far below at $M_s$ -190 K. Fitting of scaling exponent shown at $M_s$ from Eq. 3.

C. Critical exponents and universality classes

Having established the continuous nature and the expected scaling in domain size distribution, we now discuss how the related scaling laws and associated critical exponents compare with known systems that exhibit critical behavior. Figure 6 shows three critical exponents for various critical systems, comparing literature data (in black) with own results from the MD simulations (vertical blue lines)[9,17,19,22,37–41]. We note, that the exponent $\beta$ is notoriously difficult to extract from atomistic simulations given the large fluctuations originating both from the relatively small system sizes and the nature of second order phase transitions. The range reported for $\beta$ accounts for the variability associated with the various system sizes and the fitting procedure itself, see Supplementary Figures 1-2. Our values are comparable both to the single experimental example of critical martensite (with an external field), as well as to several other critical phenomena (including ferromagnetism, ferroelectricity, liquid-gas coexistence, and others shown in Supplemental Table 2), as demonstrated in Fig. 6A. The values for 65 at. % NiAl systems range from 0.45 to 0.7 and increase significantly for the highest NiAl fractions where the transformation is suppressed. The calculated values of $\tau$ for 65 at. % NiAl in Fig. 6B range from 1.3 to 1.6 are also comparable to other critical transformations: experimental piezoelectric force microscopy in a relaxor ferroelectric[37], as well as spallation[38] and granular materials[22] from MD (see Supplemental Table 3). Here, the use of the direct atomic data (rather than overall system response) for the fitting of the $\tau$ exponent makes it much less sensitive to the small size effects as compared to $\beta$. Finally, the critical exponent $\gamma$ governs the temperature dependence of the derivative of the order parameter with respect to applied field as the critical point is approached thermally:

$$\chi = \frac{\partial \varepsilon_T}{\partial \varepsilon_C} = (T - M_s)^{-\gamma} \qquad (4)$$

where $\varepsilon_T$ is the transformation strain, $\varepsilon_C$ is the coherency strain from the second phase, $T$ is temperature, and $M_s$ is the martensite start (critical) temperature. In our case, the gradient of transformation strain with respect to the applied strain field is computed using multiple values of the volume fraction of the second phase; fits are shown in Supplemental Fig. 6. The resulting value of $\gamma$, 0.85, calculated using all volume fractions from 60-75 at. % NiAl also corresponds reasonably well to existing critical behavior, shown in Fig. 6C (see also Supplemental Table 4) with a range from fitting, with the same discrepancy and small size fluctuations as discussed above for $\beta$.

The values near $\beta \sim 0.5$ and $\gamma$ close to 1 indicate that our system may fit within the mean-field universality class. Additional work will be required to definitively assess the universality class that these transformations belong to.

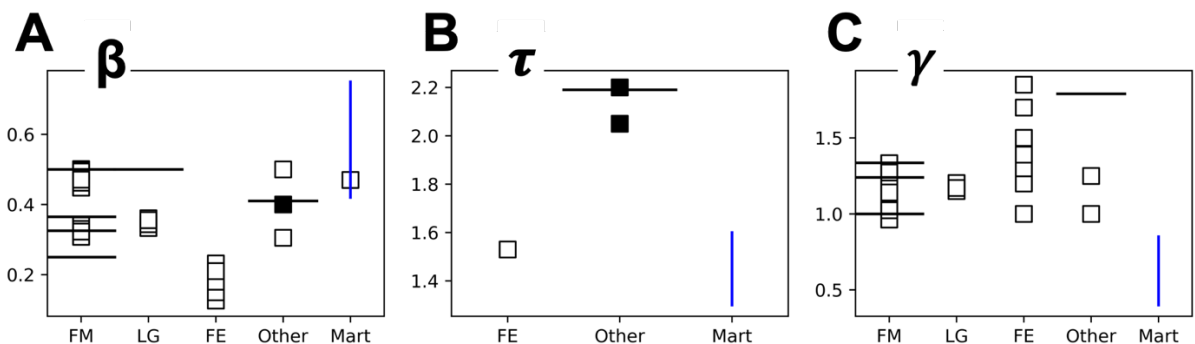

**Fig. 6.** Comparison of critical exponents. Data shown for (A) $\beta$, (B) $\tau$, and (C) $\gamma$ in the current martensitic study, with 65 at. % NiAl for $\beta$ and $\tau$, (Mart.) and well-known transformations: ferromagnetism (FM), liquid-gas coexistence (LG), ferroelectricity (FE), and other behavior. Horizontal lines represent theoretical results, open symbols experimental, and filled symbols MD (vertical line ranges for martensite). All values are described within Supplemental Tables 2-4.

## V. Conclusion and outlook

In summary, we have shown that stress fields originating from a tailored second phase can change the nature of martensitic transformation in metallic alloys from first to second order. We have demonstrated the conversion of a first order martensitic transformation to second order through MD simulations contrasting the behavior of a bulk alloy to nanolaminates with internal, epitaxial stress from a second phase. We have shown this tuning through critical exponents (which cannot be defined for the bulk alloy) related to the evolution of the transformation (order parameter), martensite domain cluster number distributions, and a "strain-susceptibility". Importantly, our simulations show that the transformation can be described via scaling laws with exponent consistent with the mean-field universality class. Further, and perhaps most clearly, is the change from abrupt to continuous transition between the bulk and nanolaminate, with a corresponding reduction of the thermal hysteresis to zero. Characterization of the bulk system matching Fig. 3 and Fig. 5 highlights the clear difference in behavior between the first and second order systems (Supplemental Fig. 7). In addition, irregular austenite/martensite interfaces persist within the critical nanolaminates throughout much of this continuous transition in stark contrast to the bulk alloy.

Beyond the application of external fields discussed above, changes in the nature of the transformation have been demonstrated in ferroelectrics going from bulk to thin films[42] or via randomized composition[43], while martensitic transformations in metallic alloys are considered to be universally first order, with few examples of compositions with continuous behavior[13–16]. While there has been significant investigation of martensitic thin films and microstructures arising from constraints[44], it has not focused on using these constraints to modify the order of the transformation. Interestingly, recent work has shown intriguing "higher order" characteristics in certain gum metals containing nanoscale variations in composition[23,24,45]. Ref. 23 showed both continuous stress-induced martensitic transformation due to mechanical confinement and a minimum in stiffness near the critical temperature. These observations are consistent with our theoretical findings: Fig. 2 shows continuous thermal transformation and Fig. 7 shows stiffness as a function of temperature for high volume fractions of NiAl strikingly similar to the aforementioned experimental results, with large, v-shaped dips in stiffness near the transformation temperature. Each sample used to evaluate stiffness was taken from the cooling simulations, equilibrated for 50 ps, and strained biaxially to 1% at a rate of $2 \cdot 10^8$ s$^{-1}$ parallel to the laminate interface. A linear fit was taken from the resulting stress-strain data to obtain the biaxial modulus.

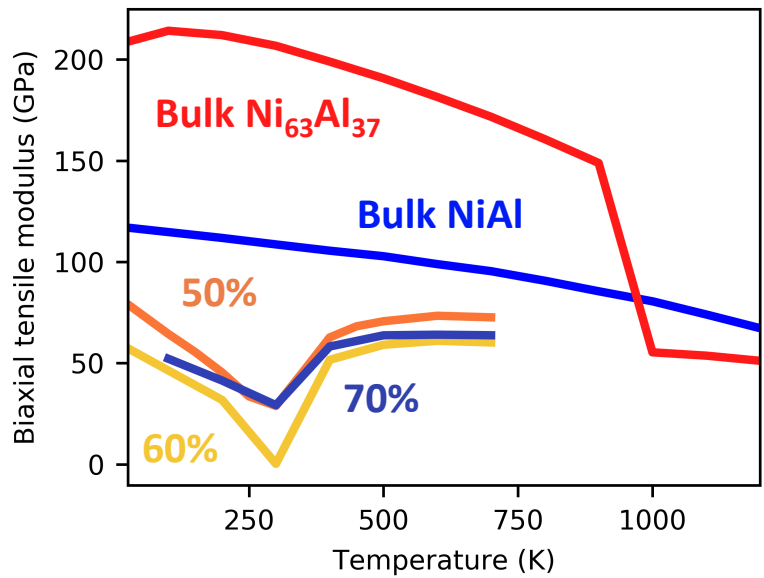

**Figure 7:** Stiffness in biaxial tension as a function of temperature for bulk phases and nanolaminates

Finally, we note that FELE has previously been used to create exciting new properties and functionalities in other systems; for example, ferroelectrics with negative capacitance[6], ultra-high piezoelectricity[11], and enhanced ferroelectricity through the related area of strain engineering[46,47] have been demonstrated. The FELE approach described to transform a martensitic transition into a second order is rather general and could be applied to any alloy where an appropriate second phase can be identified. This could be used to develop new martensitic materials with ultra-low hysteresis for sensing and actuation applications and generally enable tighter control for those applications with tunable and continuous transformations in the desired temperature range and with the needed strain response. This includes potentially extending the experimentally demonstrated, externally loaded Fe-Pd systems with continuous transformation to more feasible engineering applications with material-level tuned critical martensite. We anticipate that a coherent second phase could be incorporated via layered deposition (coherent metallic superlattices have been demonstrated using magnetron sputtering[48] and molecular beam epitaxy[49]) or by introducing coherent precipitates via traditional metallurgical processing[50].

## Supplemental Material

Supplemental Figures 1-7 show the details of fitting critical exponents for all martensitic systems studied.

Supplemental Table 1 shows the details of systems sizes shown in supplemental comparisons.

Supplemental Tables 2-4 show reference data critical exponents for many other phase transitions.

Supplemental Movies 1-2 contrasts the atomistic process of martensitic transformation for the 65 at. % NiAl layered metamaterial and bulk $Ni_{63}Al_{37}$ alloy, respectively.

## Acknowledgements

This work was supported by the United States Department of Energy Basic Energy Sciences (DoE-BES) program under Program No. DE-FG02-07ER46399 (Program Manager John Vetrano). This work was performed in part under the auspices of the U.S. Department of Energy by Lawrence Livermore National Laboratory under Contract DE-AC52-07NA27344. Computational resources from nanoHUB and Purdue University are gratefully acknowledged.

43. Lemanov, V. V., Smirnova, E. P., Syrnikov, P. P. & Tarakanov, E. A. Phase transitions and glasslike behavior in Sr(1-x)Ba(x)TiO3. *Phys. Rev. B* **54**, 3151–3157 (1996).

44. Roytburd, A. L., Kim, T. S., Su, Q., Slutsker, J. & Wuttig, M. Martensitic transformation in constrained films. *Acta Mater* **46**, 5095–5107 (1998).

45. Zhu, J., Gao, Y., Wang, D., Zhang, T.-Y. & Wang, Y. Taming martensitic transformation via concentration modulation at nanoscale. *Acta Mater.* **130**, 196–207 (2017).

46. Haeni, J. H. *et al.* Room-temperature ferroelectricity in strained $SrTiO_3$. *Nature* **430**, 758–761 (2004).

47. Choi, K. J. *et al.* Enhancement of Ferroelectricity in Strained $BaTiO_3$ Thin Films. *Science* **306**, 1005–1009 (2004).

48. Wei, Q. M., Liu, X.-Y. & Misra, A. Observation of continuous and reversible bcc–fcc phase transformation in Ag/V multilayers. *Appl. Phys. Lett.* **98**, 111907 (2011).

49. Park, S. *et al.* Tunable magnetic anisotropy of ultrathin Co layers. *Appl Phys Lett* **86**, 2–4 (2005).

50. Khalil-Allafi, J., Dlouhy, A. & Eggeler, G. $Ni_4Ti_3$-precipitation during aging of NiTi shape memory alloys and its influence on martensitic phase transformations. *Acta Mater.* **50**, 4255–4274 (2002).
15

# Supplemental material:

# Uncharacteristic second order martensitic transformation in metals via epitaxial stress fields

**Authors:** Samuel Temple Reeve[1], Karthik Guda Vishnu[2], and Alejandro Strachan[2]

**Affiliations:**

[1]Materials Science Division, Lawrence Livermore National Laboratory,
7000 East Ave, Livermore, CA 94550

[2]School of Materials Engineering and Birck Nanotechnology Center, Purdue University, West Lafayette, Indiana 47906 USA.

*Correspondence to: strachan@purdue.edu

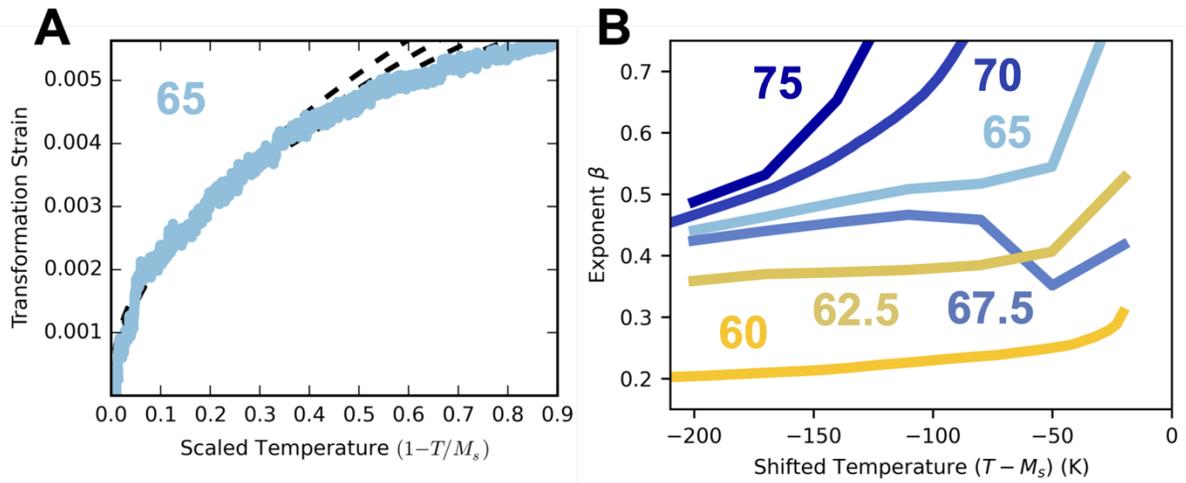

**Supplemental Figure 1:** Calculation of exponent $\beta$. (A) fitting example for 65 at. % NiAl and (B) exponents as a function of temperature.

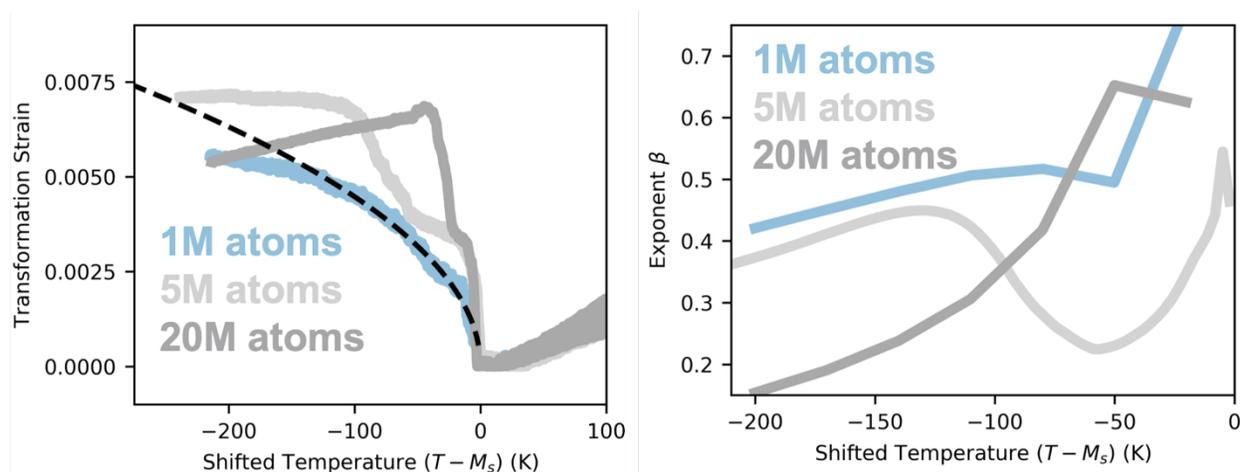

**Supplemental Figure 2:** Calculation of exponent $\beta$ for multiple system sizes of 65 at. % NiAl. (A) fitting for three total volumes (B) exponents as a function of temperature for three total volumes.

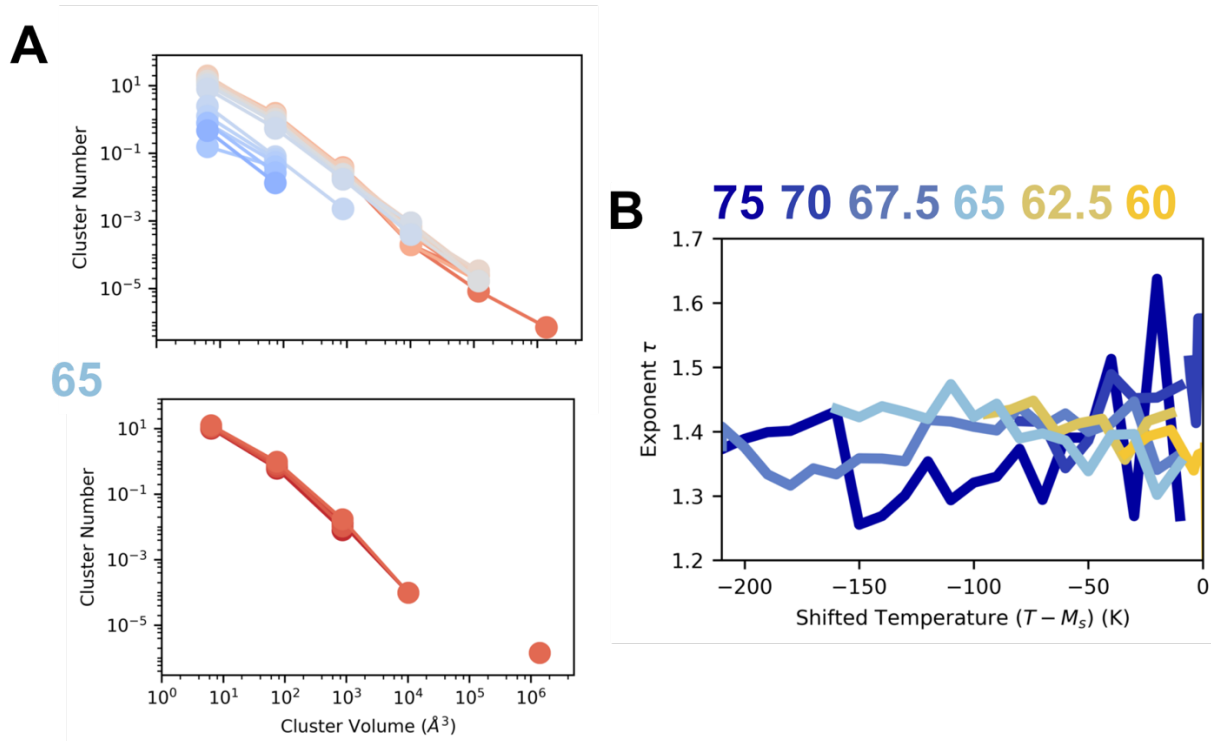

**Supplemental Figure 3:** Calculation of exponent $\tau$. (A) cluster distribution at and below $M_s$ for 65 at. % NiAl (upper) and far below $M_s$ for 65 at. % NiAl (lower) and (B) exponents as a function of temperature below the transition.

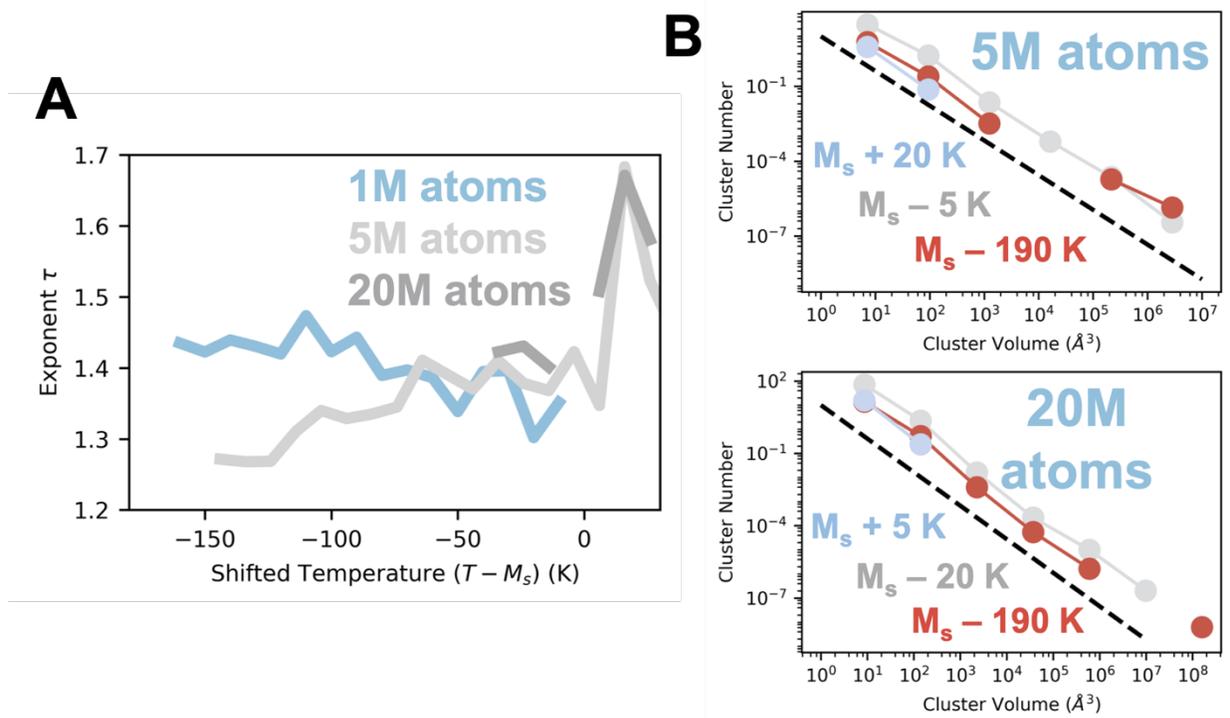

**Supplemental Figure 4:** Calculation of exponent $\tau$ for multiple system sizes of 65 at. % NiAl.. (A) exponents as a function of temperature and (B) cluster distribution near $M_s$ for 5 (top) and 20 (bottom) million atoms.

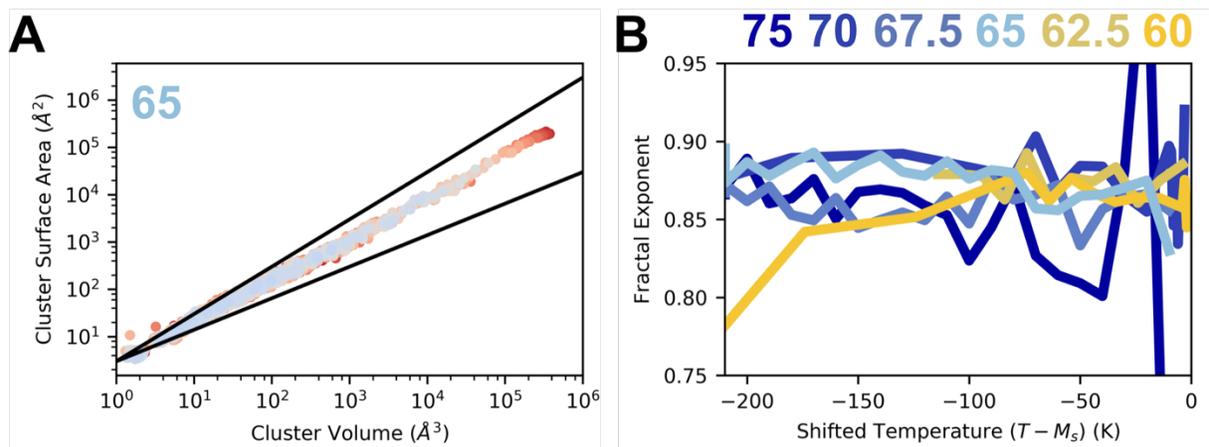

**Supplemental Figure 5:** Calculation of fractal exponent. (A) fitting example for 65 at. % NiAl, with guides of slope 1 and 2/3 (bulk), and (B) exponents as a function of temperature.

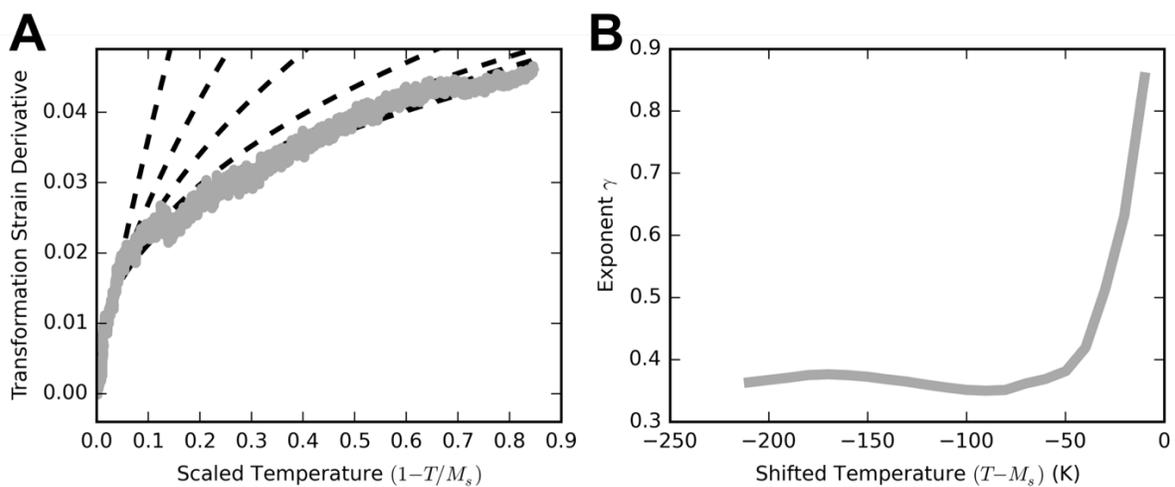

**Supplemental Figure 6:** Calculation of exponent $\gamma$. (A) fitting and (B) exponent as a function of temperature.

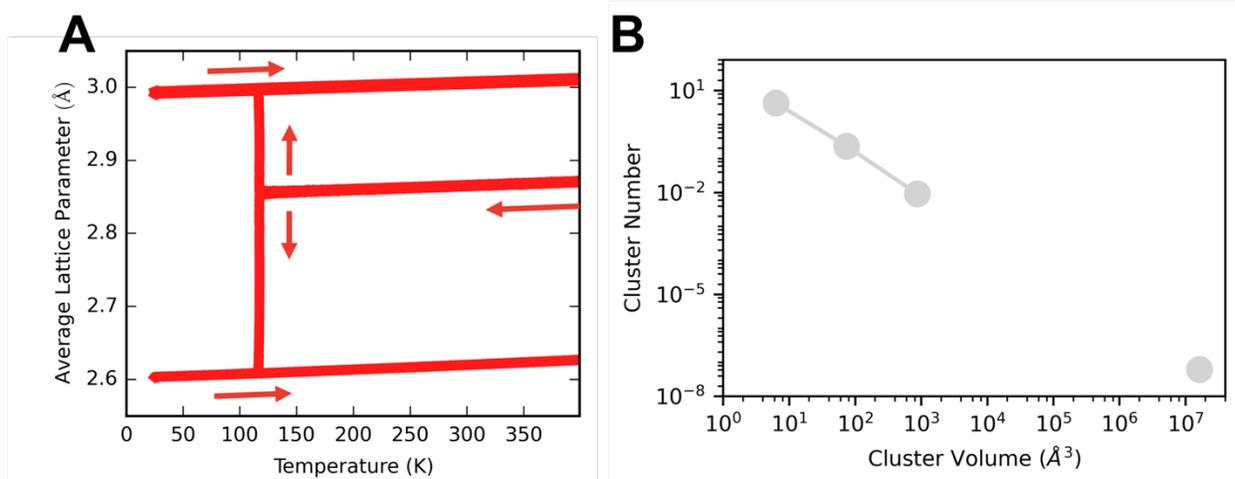

**Supplemental Figure 7:** Bulk $Ni_{63}Al_{37}$ characteristics. (A) cooling and heating cycle and (B) cluster distribution at $M_s$ with near complete transformation.

**Supplemental Table 1:** MD system sizes

| Atoms | Unit cells per direction | Initial length per direction |
|---|---|---|
| 1,024,000 | 80 | 23.2 nm |
| 5,488,000 | 140 | 40.6 nm |
| 20,155,392 | 216 | 62.6 nm |

**Supplemental Table 2:** Beta critical exponents from literature

| Behavior | Material/Theory | Exponent | Reference |
|---|---|---|---|
| Martensite (Mart) | Fe-31.2%Pd | 0.47 | 17 |
| Ferromagnetic (FM) | Mean field | 0.5 | 9 |
| | 3D Heisenberg | 0.365 | |
| | 3D Ising | 0.325 | |
| | Tri-critical mean field | 0.25 | |
| | Iron | 0.34 | |
| | Nickel | 0.33 | |
| | $Nd_{0.5}Sr_{0.5}MnO_3$ | 0.323 | 39 |
| | $Pr_{0.5}Sr_{0.5}MnO_3$ | 0.448 | |
| | $Ni_{50}Mn_{35}Sn_{15}$ | 0.501 | 40 |
| Liquid-Gas Coexistence (LG) | van der Waals | 0.5 | 9 |
| | $CO_2$ | 0.333 | |
| | Xe | 0.345 | |
| | He | 0.361 | |
| Ferroelectric (FE) | $Sr_{0.61-x}Ce_xBa_{0.39}Nb_2O_6$ | 0.14 | 37 |
| Order-Disorder | CuZn | 0.305 | 9 |
| Mott | $(Cr_{0.989}V_{0.011})_2O_3$ | 0.5 | 19 |
| Spallation | Ta (MD) | 0.4 | 38 |
| Percolation | 3D | 0.41 | 41 |

**Supplemental Table 3:** Tau critical exponents from literature

| Behavior | Material/Theory | Exponent | Reference |
|---|---|---|---|
| Ferroelectric (FE) | $Sr_{0.61-x}Ce_xBa_{0.39}Nb_2O_6$ (2D measurement) | 1.53 | 37 |
| Spallation | Ta (MD) | 2.2 | 38 |
| Granular material | (MD) | 2.06 | 22 |
| Percolation | 3D | 2.18 | 41 |

**Supplemental Table 4:** Gamma critical exponents from literature

| Behavior | Material/Theory | Exponent | Reference |
|---|---|---|---|
| Ferromagnetic (FM) | Mean field | 1.0 | 9 |
| | 3D Heisenberg | 1.336 | |
| | 3D Ising | 1.24 | |
| | Tri-critical mean field | 1.0 | |
| | Iron | 1.333 | |
| | Nickel | 1.32 | |
| | $Nd_{0.5}Sr_{0.5}MnO_3$ | 1.201 | 39 |
| | $Pr_{0.5}Sr_{0.5}MnO_3$ | 1.334 | |
| | $Ni_{50}Mn_{35}Sn_{15}$ | 0.963 | 40 |
| Liquid-Gas Coexistence (LG) | van der Waals | 1.0 | 9 |
| | $CO_2$ | 1.20 | |
| | Xe | 1.203 | |
| | He | 1.15 | |
| Ferroelectric (FE) | $Sr_{0.61-x}Ce_xBa_{0.39}Nb_2O_6$ | 1.0 | 37 |
| Order-Disorder | CuZn | 1.25 | 9 |
| Mott | $(Cr_{0.989}V_{0.011})_2O_3$ | 1.0 | 19 |
| Percolation | 3D | 1.80 | 41 |